\documentclass[aps,amsmath,amssymb,nofootinbib,twocolumn]{revtex4}
\usepackage{graphicx}
\usepackage{pgfplots}
\pgfplotsset{compat=1.17}
\usepackage{dcolumn}
\usepackage{color}
\usepackage{scrextend}
\usepackage{subfig}
\usepackage{bm}
\usepackage{amsmath,amssymb,graphicx}
\usepackage{epsfig}
\usepackage{enumerate}
\usepackage{array}
\usepackage{multirow}
\usepackage{tikz}
\usepackage{ragged2e}
\usepackage{textgreek}
\usepackage[normalem]{ulem}

\begin{document}
\title{Non-Markovian renormalization of optomechanical exceptional points}
\author{Aritra Ghosh\footnote{aritraghosh500@gmail.com} and M. Bhattacharya}
\affiliation{School of Physics and Astronomy, Rochester Institute of Technology, 84 Lomb Memorial Drive, Rochester, New York 14623, USA}
\vskip-2.8cm
\date{\today}
\vskip-0.9cm

\vspace{5mm}
\begin{abstract}
We investigate how non-Markovian mechanical dissipation affects exceptional points in linearized optomechanical systems with red-sideband drive. For a chosen non-Ohmic mechanical bath, we derive analytical conditions for the memory-renormalized exceptional point by employing a pseudomode mapping, thereby demonstrating that structured environments displace the mode coalescence away from the Markovian prediction. Crucially, we reveal that failing to account for this memory-induced shift suppresses the divergent Petermann factor by orders of magnitude, showing that accurate bath modeling is essential for the successful operation of exceptional-point-based devices whenever reservoir-induced memory is non-negligible. We finally show that non-Markovianity modifies the cavity reflection spectrum, manifesting as a shallower optomechanically-induced-transparency dip, providing therefore an experimentally-accessible signature of structured mechanical environments.
\end{abstract}

\maketitle

\section{Introduction}
The recent years have witnessed a surge in the interest in non-Hermitian quantum systems \cite{Moiseyev_2011,Bagarello_2015}, largely motivated by the developments in open quantum systems that admit nonunitary time evolutions \cite{Rotter_2009}. Of great importance is the occurrence of exceptional points where two or more eigenstates (eigenvalues plus eigenvectors) coalesce \cite{Berry_2004,Heiss_2012}. Exceptional points have not only been observed experimentally \cite{Kim_2016,Liang_2023}, but are also important for understanding quantum phase transitions \cite{Stransky_2018,Hamazaki_2021}, quantum interference \cite{Klauck_2025}, quantum entanglement \cite{Chakraborty_2019,Wang_2025}, topological aspects \cite{Bergholtz_2021}, etc., in non-Hermitian systems. Potential applications include lasing \cite{Peng_2014,Zhang_2022}, sensing \cite{Lau_2018,Wiersig_2020}, mode switching \cite{Doppler_2016}, etc. Recent studies in open quantum systems have also explored the distinction between Liouvillian and Hamiltonian exceptional points \cite{Minganti_2019,Minganti_2020,Khandelwal_2021,Abo_2024}, as well as the notion of non-Markovian exceptional points \cite{Lin_2025,Zhang_2025}. 

\vspace{2mm}

The framework of cavity optomechanics supplies a versatile platform for exploring light-matter interactions at the quantum level, enabling applications ranging from precision sensing to tests of macroscopic quantum phenomena \cite{Kippenberg_2007,Aspelmeyer_2014}, while also serving as a useful platform for studying non-Hermitian quantum mechanics \cite{Chakraborty_2019,Wang_2025,Xiong_2021,Pino_2022,Sun_2023}. While in the standard theoretical description, the mechanical oscillator is assumed to interact with a Markovian thermal environment \cite{Aspelmeyer_2014}, recent developments have demonstrated the experimental observation of strong non-Markovian character \cite{Groeblacher_2015}, significantly deviating from the Ohmic scaling of the bath spectrum. This raises the natural question as to how the non-Markovian nature of the mechanical dissipation can affect the exceptional points in optomechanical systems. Structured reservoirs introduce memory effects that cannot be captured within the conventional Markovian framework, and understanding how these non-Markovian mechanical environments modify the non-Hermitian structure of optomechanical dynamics, in particular the fate of optomechanical exceptional points, therefore constitutes an important open problem both for the fundamental theory of open quantum systems and for the interpretation of experiments operating in regimes where reservoir-induced memory is not negligible.

\vspace{2mm}

In this paper, we shall theoretically demonstrate how a chosen structured non-Markovian mechanical environment modifies the non-Hermitian structure and exceptional point of an optomechanical system. Our analysis hinges on the linearized regime to reduce the dynamics to a tractable bilinear form, alongside a red-sideband drive to facilitate the beamsplitter interaction that underlies the emergence of optomechanical exceptional points. Considering an optomechanical cavity in which the mechanical oscillator interacts with a structured environment characterized by a super-Ohmic spectral density at low frequencies, we will map the resulting non-Markovian drift dynamics onto an equivalent Markovian drift through the pseudomode embedding by introducing an auxiliary mode that captures the finite memory of the bath \cite{Lin_2025,Garraway_1997}. This construction allows the drift dynamics to be described by an enlarged set of coupled Markovian quantum Langevin equations, enabling a systematic analysis of the spectrum of the effective drift matrix. 

\vspace{2mm}

Using this framework, we shall derive analytic expressions for the exceptional points accessible for realistic parameters and show how reservoir-induced memory shifts the optomechanical exceptional point without destroying it. For realistic parameters, our analysis shows that the resulting deformation of the exceptional-point location may reach the percent level. Crucially, it is found that memory-induced corrections lead to dramatic changes in the values of the Petermann factor \cite{Berry_2003}. Moreover, by analyzing the cavity's input-output response, we shall demonstrate that the non-Markovian mechanical memory alters the interference condition responsible for optomechanically-induced transparency. As a consequence, the memory-renormalized susceptibility produces a quantitatively-shallower transparency dip in the reflection spectrum, providing a robust and directly-measurable fingerprint of the structured environment.

\vspace{2mm}

It should be emphasized that while the pseudomode construction for non-Markovian systems is a well-known approach, our motivation is to apply this framework to cavity optomechanics, in order to unravel the effect of environment-induced memory on the optomechanical exceptional point. The present work is therefore distinguished from earlier general studies of non-Markovian exceptional points \cite{Lin_2025,Zhang_2025} in three respects. It identifies a concrete optomechanical setting in which structured mechanical dissipation shifts the red-sideband exceptional point, derives both perturbative and numerically-exact conditions for this shifted degeneracy within the pseudomode-embedded drift dynamics, and connects the resulting displacement to measurable optomechanical signatures, namely, the Petermann factor and the cavity reflection spectrum.

\vspace{2mm}

The paper is organized as follows. In Sec. (\ref{model_sec}), we will introduce the optomechanical model accompanied by non-Markovian mechanical dissipation with low-frequency super-Ohmic behavior. This is followed by our presentation of the pseudomode embedding of the non-Markovian optomechanical drift dynamics within an enlarged Markovian framework in Sec. (\ref{pseudo_sec}). Sec. (\ref{NMEP_sec}) then analyzes the memory-induced corrections to the optomechanical exceptional point and its consequences in the Petermann factor. Finally, we shall discuss observable signatures of non-Markovianity in the reflection spectrum in Sec. (\ref{spec_sec}) and then briefly discuss our results in the context of generic spectral densities in Sec. (\ref{gen_sec}). The paper is concluded in Sec. (\ref{conc_sec}). Technical details of the calculations are relegated to the Appendices (\ref{appA})-(\ref{appE}). 

\section{Theoretical model}\label{model_sec}
\begin{figure}
\centering
\includegraphics[width=1\linewidth]{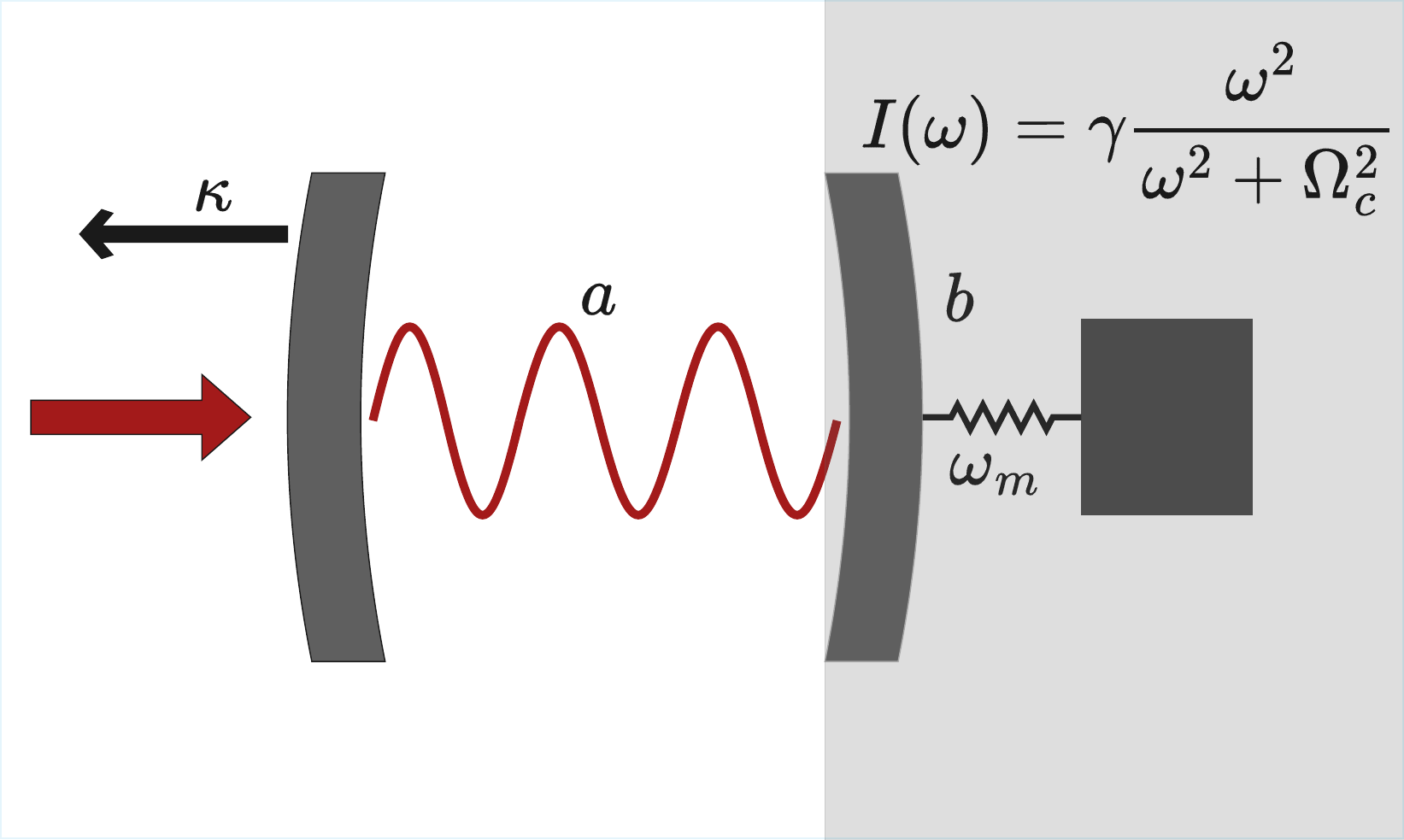}
\caption{\justifying{Schematic of an optomechanical setup where a control laser enters the cavity from the left mirror and the intracavity mode $a$ couples with the micromechanical motion of the right mirror, represented by the oscillator-mode $b$. The mechanical eigenfrequency is $\omega_m$ and $G$ is the effective (linearized) optomechanical coupling. The optical decay rate is $\kappa$, while the structured bath shaded in gray is characterized by the damping scale $\gamma$. We shall take $\omega_m \gg \kappa, G, \gamma$ in conformity with the resolved-sideband operation.}}
\label{fig1}
\end{figure}

The theoretical model consists of a linearized optomechanical system with red-sideband drive, where the cavity optical field couples to the micromechanical Brownian motion of one of the mirrors, as shown in Fig. (\ref{fig1}). The coherent part of the dynamics, i.e., the optomechanical dynamics, is described by the following Hamiltonian in a frame rotating with the control-laser frequency (setting $\hbar = 1$) \cite{Kippenberg_2007,Aspelmeyer_2014}:
\begin{equation}
H = - \Delta a^\dagger a + \omega_m b^\dagger b + G(a^\dagger b + a b^\dagger),
\end{equation}
where $\Delta$ is the detuning of the control laser from the cavity resonance, $\omega_m$ is the eigenfrequency of the mechanical oscillator, and $G$ is the linearized optomechanical coupling, proportional to the square root of the input power of the control drive, i.e., $G \propto \sqrt{P_{\rm in}}$. The above Hamiltonian is obtained from the nonlinear radiation-pressure-type optomechanical Hamiltonian by linearizing the cavity field about the control drive \cite{Kippenberg_2007,Aspelmeyer_2014}, so $(a,a^\dagger)$ are the operators describing the cavity optical fluctuations and $(b,b^\dagger)$ are their mechanical counterparts. The linearization has been supplemented by the rotating-wave approximation (in the resolved-sideband regime) near the red sideband, $\Delta \simeq -\omega_m$, thereby dropping the two-mode-squeezing terms and retaining only the beamsplitter interaction \cite{Kippenberg_2007,Aspelmeyer_2014}. By definition, $\omega_m$ denotes the calibrated baseline mechanical frequency of the effective red-sideband model, including the Bloch-Siegert-type shift associated with the counter-rotating terms being neglected in the rotating-wave approximation, while the static reactive shift generated by the mechanical bath is removed by the counterterm convention discussed in Appendix (\ref{appA}). In our analysis, we shall refer to $\omega_m$ as the bare mechanical frequency, in the restricted sense of this calibrated baseline frequency.

\vspace{2mm}

Accounting for dissipation, the coupled set of quantum Langevin equations are
\begin{eqnarray}
\dot{a} &=& i \Delta a - \frac{\kappa}{2}a -iGb + \sqrt{\kappa} a_{\rm in}(t), \label{Lang_a} \\
\dot{b} &=& -i\omega_m b -
\int_0^t K(t-\tau)b(\tau)d\tau - iGa + F(t), \label{Lang_b}
\end{eqnarray}
where $\kappa $ is the optical damping rate, and the mechanical oscillator exhibits a non-Markovian damping. The input noises \cite{Gardiner_1985} are $a_{\rm in}(t)$ and $F(t)$. The optical input fluctuations admit the Markovian correlations
\begin{equation}
\langle a_{\rm in}(t) a^\dagger_{\rm in}(t') \rangle = \delta(t-t'), \quad 
\langle a^\dagger_{\rm in}(t) a_{\rm in}(t') \rangle = 0,
\end{equation}
valid for optical frequencies at room temperature \cite{Aspelmeyer_2014}. In contrast, the mechanical oscillator exhibits non-Markovian character with dissipation characterized by the memory kernel $K(t)$. The corresponding noise correlations are (see Appendix (\ref{appA}))
\begin{eqnarray}
\langle F(t)F^\dagger(t')\rangle &=& \int_0^\infty \frac{d\omega}{2\pi} I(\omega) (n(\omega)+1)e^{-i\omega(t-t')}, \label{FDT_mech} \\
\langle F^\dagger(t)F(t')\rangle &=& \int_0^\infty
\frac{d\omega}{2\pi} I(\omega) n(\omega) e^{i\omega(t-t')}, \nonumber
\end{eqnarray} 
where $n(\omega)$ is the thermal Bose factor and $I(\omega)$ is the bath spectral function. For our purposes, the memory kernel is related to the bath spectral function as
\begin{equation}\label{K_int}
K(t)=\int_{-\infty}^{\infty}\frac{d\omega}{2\pi}\mathcal{I}(\omega)e^{-i\omega t},
\end{equation}
where 
\begin{equation}
\mathcal{I}(\omega)=I(|\omega|),
\end{equation}
which is the even extension of the positive-frequency effective bath profile. In Eq. (\ref{K_int}), the normalization of this two-sided spectrum is fixed by the kernel definition itself and any convention-dependent factor relative to the one-sided microscopic spectral density discussed in Appendix (\ref{appA}) is absorbed into the effective positive-frequency profile $I(\omega)$, and hence into the overall damping scale. It should be emphasized that this symmetric kernel is an effective representation of the mechanical bath. The microscopic derivation yields instead the one-sided kernel (see Appendix (\ref{appA})) which contains both dissipative and reactive contributions. In the present treatment, the odd-in-frequency reactive part is absorbed by introducing a counterterm in the effective system-plus-bath Hamiltonian, so we can work with the even-in-frequency kernel introduced above. Let us model the bath spectral function as
\begin{equation}\label{structured_spectral_density}
\mathcal{I}(\omega)=\gamma\frac{\omega^2}{\omega^2+\Omega_c^2},
\end{equation}
exhibiting super-Ohmic scaling for $|\omega|\lesssim\Omega_c$, saturating to a constant above the crossover scale $\Omega_c$. Here $\gamma$ denotes the overall damping scale of the structured reservoir. The associated memory kernel is
\begin{equation}\label{Kt_explicit}
K(t)=\gamma\delta(t)-\frac{\gamma\Omega_c}{2}e^{-\Omega_c|t|}.
\end{equation}
This kernel therefore contains both a local Markovian contribution and an exponentially-decaying non-Markovian contribution. The above-mentioned form of the spectral density of the bath will form the basis of our analysis. In particular, the spectral density in Eq. (\ref{structured_spectral_density}) should be understood as a minimal analytically-tractable model of structured mechanical dissipation that can be tackled using the pseudomode-embedding approach by introducing a single auxiliary degree of freedom.

\section{Pseudomode embedding}\label{pseudo_sec}
Let us invoke the pseudomode embedding that hinges upon the idea of transforming the non-Markovian equations into an effective set of coupled Markovian equations through the introduction of an auxiliary mode $c$ \cite{Garraway_1997}. This leads to the equations (see Appendix (\ref{appB}))
\begin{eqnarray}
\dot{a} &=& i\Delta a-\frac{\kappa}{2}a-iGb+\sqrt{\kappa} a_{\rm in}(t), \label{a_extended}\\
\dot{b} &=& -i\omega_m b-\frac{\gamma}{2}b-iGa-g_c c+\sqrt{\gamma} \xi(t), \label{b_extended}\\
\dot{c} &=& -\Omega_c c- g_c b+\sqrt{2\Omega_c} \xi(t) \label{c_extended},
\end{eqnarray}
where $\xi(t)$ is an effective Markovian noise that the modes $b$ and $c$ couple to, and whose correlations are given by
\begin{eqnarray}
\langle \xi(t)\xi^\dagger(t')\rangle &=& (n_{\rm th}+1)\delta(t-t'), \\ 
\langle \xi^\dagger(t)\xi(t')\rangle &=& n_{\rm th}\delta(t-t'), \nonumber
\end{eqnarray} for a constant occupation number $n_{\rm th}$. We have also introduced the parameter $g_c = \sqrt{\gamma\Omega_c/2}$. Eliminating the auxiliary mode $c(t)$ reproduces the exponential memory kernel of the effective non-Markovian mechanical dynamics, up to an initial transient that decays as $e^{-\Omega_c t}$. In the long-time stationary regime, therefore, the drift part of the evolution is embedded exactly, leading to the drift matrix
\begin{equation}\label{general_drift_matrix}
M =
\begin{pmatrix}
i\Delta-\frac{\kappa}{2} & -iG & 0
\\
-iG & -\left(i\omega_m+\frac{\gamma}{2}\right) & -g_c
\\
0 & -g_c & -\Omega_c
\end{pmatrix},
\end{equation} 
obtained by casting the quantum Langevin equations in the matrix form. The eigenproblem of $M$ is the natural pathway to non-Markovian exceptional points. The cubic characteristic equation reads
\begin{eqnarray}
(\lambda+\Omega_c)
\left[(\lambda+i\omega_m+\tfrac{\gamma}{2})
(\lambda-i\Delta+\tfrac{\kappa}{2}) + G^2\right]\\
- g_c^2(\lambda-i\Delta+\tfrac{\kappa}{2})=0, \nonumber
\end{eqnarray} and exceptional-point degeneracies can be obtained from here. It can be verified that the isolated mechanics-plus-pseudomode block does not exhibit an exceptional point for physical parameter values of interest; that requires $\Omega_c=\frac{\gamma}{2}$ and $\omega_m=\gamma$, which is not of physical interest because $\omega_m \gg \gamma$. The relevant exceptional-point physics therefore arises in the optomechanical sector that acquires modifications due to the non-Markovian mechanical motion.

\section{Non-Markovian exceptional points}\label{NMEP_sec}

\subsection{Non-Markovian corrections at leading order}
It is possible to reduce the matrix $M$ into an effective $2 \times 2$ form for the optomechanical block via Schur-complement reduction, giving (see Appendix (\ref{appC}))
\begin{equation}
M_{{\rm eff}}(\lambda) =
\begin{pmatrix}
i\Delta-\frac{\kappa}{2} & -iG \\
-iG &
-(i\omega_m+\frac{\gamma}{2}) + \Sigma(\lambda)
\end{pmatrix},
\end{equation} where $\Sigma(\lambda)$ is an effective mechanical self-energy
\begin{equation}\label{self_energy_model}
\Sigma(\lambda) = \frac{g_c^2}{\Omega_c+\lambda},
\end{equation}
with $\lambda$ being an eigenvalue of $M$. It may be noted that while the $3\times3$ drift matrix $M$ in Eq. (\ref{general_drift_matrix}) is an eigenvalue-independent matrix, the reduced $2\times2$ matrix $M_{\rm eff}(\lambda)$ represents a nonlinear eigenvalue problem for the optomechanical sector, with the memory of the eliminated pseudomode encoded in the self-energy $\Sigma(\lambda)$. This self-energy represents only the residual nonlocal memory contribution, because the local Markovian part of the kernel has already been included explicitly through the term $-\gamma/2$ in the mechanical drift. At the leading order, the self-energy can be evaluated near the bare mechanical pole $\lambda \simeq -i\omega_m$, allowing us to write 
\begin{equation}
\Sigma(-i\omega_m) =
\frac{g_c^2}{\Omega_c-i\omega_m},
\end{equation}
and this directly leads to the renormalized mechanical parameters
\begin{eqnarray}
\omega_{m,{\rm eff}} &\approx& \omega_m \left[1 - \frac{\gamma \Omega_c}{2(\Omega_c^2 + \omega_m^2)}\right], \\
\gamma_{\rm eff} &\approx& \gamma \left[1 - \frac{\Omega_c^2}{\Omega_c^2 + \omega_m^2}\right].
\end{eqnarray}
Notably, $\gamma_{\rm eff}$ is the local resonant linewidth inferred from the dressed mechanical response, and is therefore distinct from the bath scale $\gamma$. In the absence of memory effects, the standard optomechanical drift matrix, lacking $\Sigma(\lambda) $, admits an exceptional point if (for $\kappa > \gamma$)
\begin{equation}
\Delta^{(0)}_{\rm EP} = - \omega_m, \quad \quad G^{(0)}_{\rm EP} = \frac{\kappa - \gamma}{4},
\end{equation}
where the superscript `0' denotes that these indicate the Markovian exceptional point, which corresponds to that from the conventional self-energy-free optomechanical model with only local mechanical damping $\gamma$. The inclusion of memory effects via the self-energy now leads to an exceptional point at
\begin{eqnarray}
\Delta_{\rm EP} &\approx& \Delta^{(0)}_{\rm EP} + \delta \Delta_{\rm EP}, \\
G_{\rm EP} &\approx& G^{(0)}_{\rm EP} + \delta G_{\rm EP},
\end{eqnarray}
where
\begin{equation}\label{shifts_explicit}
\delta \Delta_{\rm EP} = \frac{\omega_m \gamma \Omega_c}{2(\Omega_c^2 + \omega_m^2)}, \quad \quad \delta G_{\rm EP} = \frac{\gamma \Omega_c^2}{4(\Omega_c^2 + \omega_m^2)}.
\end{equation}
These expressions are leading-order estimates obtained by evaluating the self-energy near the bare mechanical pole at $\lambda\simeq -i\omega_m$. They are valid when the memory-induced correction is perturbative and slowly varying over the relevant optomechanical window, namely, when $|\Sigma(-i\omega_m)|\ll\omega_m$ and $|\Sigma'(-i\omega_m)|\ll1$. For representative parameters, this is justified by $\gamma\ll\omega_m$. Both the corrections scale linearly with the structured-bath damping scale $\gamma$, implying that stronger mechanical coupling to the structured reservoir can enhance the visibility of non-Markovian effects on the exceptional point. Experiments on membrane-based optomechanical platforms have reported MHz-scale mechanical resonators coupled to structured non-Markovian environments. In particular, in \cite{Groeblacher_2015}, the authors measured a micromechanical mode at $\sim 914~{\rm kHz}$ interacting with a non-Markovian bath. Motivated by these scales, let us consider the model parameters to be $\omega_m/2\pi = 1~{\rm MHz}$, $\kappa/2\pi = 0.2~{\rm MHz}$, $\gamma/2\pi = 5~{\rm kHz}$, and $\Omega_c/2\pi = 1~{\rm MHz}$. For these values, the Markovian theory predicts an exceptional point at $G^{(0)}_{\rm EP}/2\pi \approx 48.75~{\rm kHz}$, while the leading non-Markovian correction yields $\delta G_{\rm EP}/2\pi \approx 0.63~{\rm kHz}$. Although this correction is small compared with the cavity linewidth, it corresponds to a relative shift of approximately $1.3\%$ in the exceptional-point location and may be resolvable through precision fitting of the optomechanical response as the system parameters are tuned across the degeneracy. 

\subsection{Numerically-exact exceptional points}
In order to determine the fate of the optomechanical exceptional point in the full non-Markovian problem, not just at the leading order, it is useful to work directly with the cubic characteristic polynomial $p(\lambda)$ of the $3\times 3$ drift matrix $M$. Explicitly, 
\begin{eqnarray}
p(\lambda) &=& - g_c^2(\lambda-i\Delta+\tfrac{\kappa}{2}) \\
&&+(\lambda+\Omega_c)
\left[(\lambda+i\omega_m+\tfrac{\gamma}{2})
(\lambda-i\Delta+\tfrac{\kappa}{2}) + G^2\right], \nonumber
\end{eqnarray}
with $g_c^2=\gamma\Omega_c/2$. A second-order eigenvalue coalescence of the cubic occurs when
\begin{equation}\label{double_root_condition_generic}
p(\lambda_{\rm EP})=0,
\quad \quad
p'(\lambda_{\rm EP})=0,
\quad \quad
p''(\lambda_{\rm EP})\neq 0,
\end{equation}
which is an exceptional point when the matrix also becomes defective; for the physical solution selected below, this defectiveness is verified by the coalescence of eigenvectors and the divergence of the Petermann factor in Sec. (\ref{Petermann_sec}). Introducing the factors
\begin{eqnarray}
f(\lambda)&=&\lambda+i\omega_m+\frac{\gamma}{2}, \label{f_def} \\
g(\lambda)&=&\lambda+\Omega_c, \label{g_def}\\
h(\lambda)&=&g(\lambda)f(\lambda)-g_c^2, \label{h_def}
\end{eqnarray}
one may solve the double-root conditions to yield (see Appendix (\ref{appD}))
\begin{eqnarray}
G_{\rm EP}^2(\lambda_{\rm EP})&=&\frac{h(\lambda_{\rm EP})^2}{g(\lambda_{\rm EP})^2+g_c^2}, \label{G_EP_general} \\
\Delta_{\rm EP}(\lambda_{\rm EP})&=&
-i\left[\lambda_{\rm EP}+\frac{\kappa}{2}
+\frac{g(\lambda_{\rm EP}) h(\lambda_{\rm EP})}{g(\lambda_{\rm EP})^2+g_c^2}
\right], \label{Delta_EP_general}
\end{eqnarray}
where the physical exceptional point is obtained by selecting $\lambda_{\rm EP}$ for which $\Delta_{\rm EP}(\lambda_{\rm EP})$ and  $G_{\rm EP}^2(\lambda_{\rm EP})$ are both real, and $-\Delta_{\rm EP}>0$ and  $G_{\rm EP}^2>0$. For the representative parameter set $\omega_m/2\pi=1~{\rm MHz}$, $\kappa/2\pi=0.2~{\rm MHz}$, $\gamma/2\pi=5~{\rm kHz}$, and
 $\Omega_c/2\pi=1~{\rm MHz}$, one finds
\begin{equation}
\frac{G_{\rm EP}}{2\pi}\approx 49.38~{\rm kHz},
\quad \quad
\frac{\Delta_{\rm EP}}{2\pi}\approx -998.69~{\rm kHz},
\end{equation}
for which the coalesced eigenvalue reads
\begin{equation}
\frac{\lambda_{\rm EP}}{2\pi}
\approx
-50.62~{\rm kHz}
-i(998.72~{\rm kHz}).
\end{equation}
Note that this is a second-order exceptional point that marks the coalescence of the optomechanical hybrid modes, while the pseudomode eigenvalue remains well separated at this point and reads $\lambda_3/2\pi \approx -1001.25~{\rm kHz} -i(1.25~{\rm kHz}) $.

\vspace{2mm}

\begin{figure}
\centering
\includegraphics[width=\linewidth]{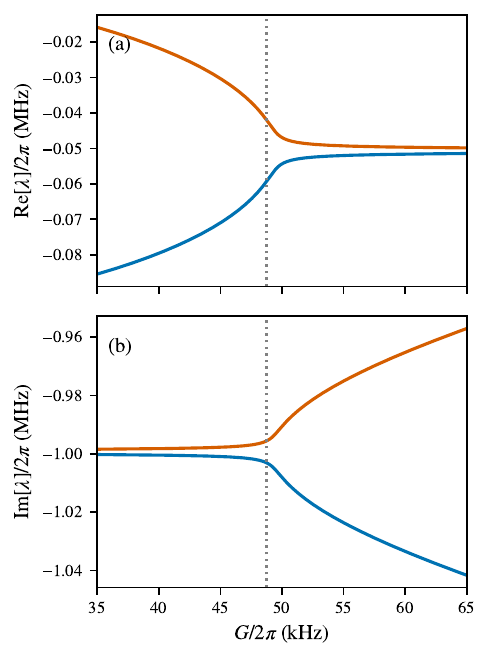}
\caption{\justifying{(a) Real part ${\rm Re}[\lambda]/2\pi$ and (b) imaginary part ${\rm Im}[\lambda]/2\pi$ of the optomechanical eigenvalues in the presence of non-Markovian effects, demonstrating avoided crossing when evaluated at the Markovian resonance $\Delta^{(0)}_{\rm EP} = -\omega_m$. The vertical dotted line indicates the location of the conventional Markovian exceptional point at $G^{(0)}_{\rm EP} = (\kappa - \gamma)/4 \approx 48.75~{\rm kHz}$. Because the system is tuned to the bare mechanical frequency rather than the memory-renormalized detuning $\Delta_{{\rm EP}} \neq \Delta^{(0)}_{\rm EP} $, the modes fail to coalesce. The model parameters are $\omega_m/2\pi = 1~{\rm MHz}$, $\kappa/2\pi = 0.2~{\rm MHz}$, $\gamma/2\pi = 5~{\rm kHz}$, and $\Omega_c/2\pi = 1~{\rm MHz}$.}}
\label{fig2}
\end{figure}

\begin{figure}
\centering
\includegraphics[width=0.9\linewidth]{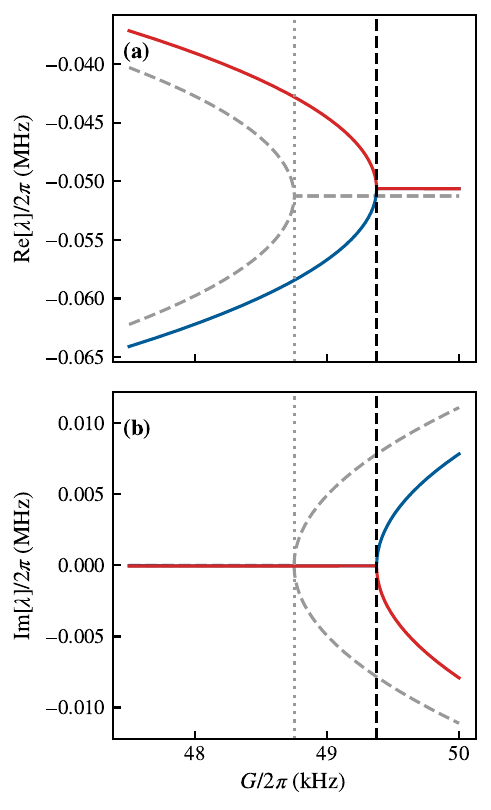}
\caption{\justifying{(a) Real part ${\rm Re}[\lambda]/2\pi$ and (b) imaginary part ${\rm Im}[\lambda]/2\pi$ of the optomechanical eigenvalues as a function of the optomechanical coupling rate $G/2\pi$ for the model parameters $\omega_m/2\pi = 1~{\rm MHz}$, $\kappa/2\pi = 0.2~{\rm MHz}$, $\gamma/2\pi = 5~{\rm kHz}$, and $\Omega_c/2\pi = 1~{\rm MHz}$. The dashed gray curves represent the bare Markovian optomechanical modes, evaluated exactly at the red-sideband resonance $\Delta^{(0)}_{\rm EP} = -\omega_m$. The solid colored curves depict the optomechanical eigenvalues of the full non-Markovian system. To observe the true non-Markovian coalescence, these eigenvalues are evaluated at the numerically-determined shifted detuning $\Delta_{\rm EP}/2\pi \approx -998.69~{\rm kHz}$. The Markovian exceptional point is indicated by the vertical dotted gray line at $G = (\kappa - \gamma)/4$, and the memory-renormalized exceptional point is marked by the vertical dashed black line.}}
\label{fig3}
\end{figure}

The effect of non-Markovianity is thus to displace the exceptional-point location, i.e., the parameters estimated using the Markovian theory no longer lead to mode coalescence when memory effects are present. Fig. (\ref{fig2}) illustrates the numerically-exact non-Markovian eigenvalues evaluated at the red-sideband resonance $\Delta = -\omega_m$. Rather than exhibiting the hallmark coalescence of an exceptional point, the hybrid optomechanical modes display a distinct avoided crossing with a persistent gap. This level repulsion visually confirms that the memory effects originating from a structured bath fundamentally shift the true exceptional point away from its original Markovian coordinates. 
However, non-Markovianity does not destroy the optomechanical exceptional point. To explicitly verify this, we have numerically evaluated the eigenvalues of the full $3 \times 3$ drift matrix $M$ across the degeneracy. As illustrated in Fig. (\ref{fig3}), the inclusion of the mechanical pseudomode shifts the exceptional point to a higher coupling strength. While our perturbative formulas provide a leading-order approximation, locating the true mathematical singularity requires evaluating the  $3 \times 3$ system at the numerically-determined non-Markovian exceptional-point detuning  $(\Delta_{\rm EP}/2\pi \approx -998.69~{\rm kHz})$. At this precise position in the parameter space, one can observe a coalescence of both the real and imaginary parts of the optomechanical eigenvalues at $G_{\rm EP}/2\pi = 49.38~{\rm kHz}$, thereby confirming that an exceptional point exists, albeit shifted from its Markovian location in the parameter space. 
It may be emphasized that rather than having an exact $\mathcal{PT}$-symmetry, the optomechanical problem discussed here belongs to a passive non-Hermitian setting (see for example, \cite{Qiu_2026}). Non-Markovianity as considered in this work does not change this but only shifts the exceptional point through the mechanical self-energy.

\subsection{Petermann factor}\label{Petermann_sec}
While the distinction between the Markovian and non-Markovian exceptional-point locations appears only within 1-2\% of the optomechanical coupling for our chosen bare parameters, a strong distinction between them can be obtained by analyzing the Petermann factor, sensitive to small displacements of the exceptional-point location. Let us consider the right and left eigenvectors of $M $ as defined by
\begin{equation}
M |R_j\rangle = \lambda_j |R_j\rangle, \quad \quad M^\dagger |L_j\rangle  = \lambda_j^* |L_j\rangle,
\label{RL_def}
\end{equation}
where $\lambda_j $ are the complex eigenvalues. For each mode $j $, the Petermann factor is defined as
\begin{equation}
K_j = \frac{\langle L_j|L_j\rangle\langle R_j|R_j\rangle}
{|\langle L_j|R_j\rangle|^2}.
\label{Petermann_def}
\end{equation}
This quantity is invariant under arbitrary rescalings of the left and right eigenvectors and reduces to unity for a normal-mode basis. In contrast, $K_j>1$ signals nonorthogonality, while $K_j\to\infty$ indicates an exceptional point, consistent with self-orthogonality of the coalescing eigenvectors \cite{Heiss_2012}. Notably, the divergence of the Petermann factor reflecting self-orthogonality of the coalescing eigenvectors is directly associated with enhanced noise sensitivity in non-Hermitian systems \cite{Berry_2003}. 

\vspace{2mm}

For our representative parameters, evaluating the  $3\times 3 $ drift matrix at the numerically-determined non-Markovian exceptional-point coordinates\footnote{The quoted exceptional-point coordinates in the text are rounded to two decimal places for readability, while the evaluation of the Petermann factors has been performed using the full-precision numerical solution of $p(\lambda_{\rm EP})=p'(\lambda_{\rm EP})=0$, giving $\Delta_{\rm EP}/2\pi = -998.6850325947269~{\rm kHz}$ and $G_{\rm EP}/2\pi = 49.37501085439698~{\rm kHz}$. Because Petermann factors are extremely sensitive to detuning from the exact exceptional point, the rounded in-text values should be understood as display values only.}, i.e., $\Delta_{\rm EP}/2\pi = - 998.69~{\rm kHz}$ and $G_{\rm EP}/2\pi = 49.38~{\rm kHz}$, one finds that the two coalescing Petermann factors become numerically singular, as in $K_+ \approx K_- \sim 10^{14}$, the finite value here being set only by numerical precision. This is the expected signature of eigenvector self-orthogonality at an exceptional point. At the same parameter values, the third eigenmode, corresponding predominantly to the pseudomode branch, remains almost orthogonal: $K_3 \approx 1.0025 $, confirming that the exceptional-point physics is confined to the optomechanical sector and does not involve the pseudomode. 

\vspace{2mm}

The contrast with the bare Markovian calibration is particularly striking. If the physical system is tuned instead to the Markovian exceptional-point coordinates, i.e., $\Delta^{(0)}_{\rm EP}/2\pi = - 1000~{\rm kHz}$ and $G^{(0)}_{\rm EP}/2\pi = 48.75~{\rm kHz}$, while the true dynamics remains non-Markovian, the Petermann factors become $K_{\pm} \approx 27.6$, while $K_3 \approx 1.0025$, as before. A parameter mismatch of only $\delta G_{\rm EP}/2\pi \approx 0.63~{\rm kHz}$ and $\delta\Delta_{\rm EP}/2\pi \approx 1.31~{\rm kHz}$, obtained from the numerically-exact non-Markovian exceptional-point coordinates (to be compared with the perturbative estimate $\delta\Delta_{\rm EP}/2\pi \approx 1.25~{\rm kHz}$), therefore suppresses the hybrid-mode nonorthogonality from a divergent exceptional-point response to a moderate value $K_\pm \sim \mathcal{O}(10)$, demonstrating that the memory-induced shift of the exceptional point, although modest at the level of the system parameters, is strongly magnified in the eigenvector structure. It should be clarified that in practice, finite uncertainty in the detuning, coupling strength, linewidths, and bath parameters regularizes the divergence. So the physically-relevant point is therefore not the absolute numerical contrast between $K_\pm\sim10^{14}$ and $K_\pm\sim30$, but rather that a small memory-induced displacement of the exceptional-point coordinates can move the system from the singular regime to a substantially-less nonorthogonal one. In other words, the Petermann factor is used here only as a diagnostic of eigenvector nonorthogonality and of the exceptional-point calibration to environmental memory.
\begin{figure}[b]
\centering
\includegraphics[width=\linewidth]{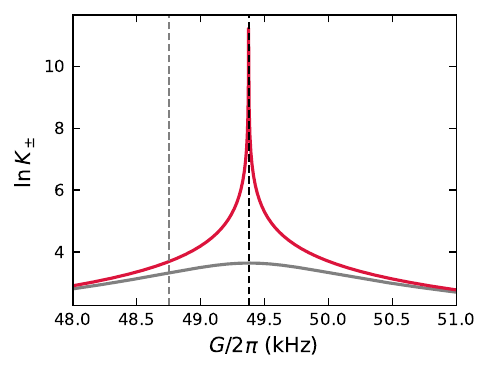}
\caption{\justifying{Divergence of the Petermann factor $K_\pm$ demonstrating the extreme sensitivity of the optomechanical exceptional point to non-Markovian mechanical dissipation. The solid red curve represents the eigenmode nonorthogonality of the full $3 \times 3$ non-Markovian system evaluated at the numerical memory-renormalized resonance $\Delta_{\rm EP}/2\pi \approx -998.69~{\rm kHz}$. It exhibits a true mathematical singularity at the non-Markovian exceptional point $G_{\rm EP}/2\pi \approx 49.38~{\rm kHz}$ (vertical black dashed line), signaling perfect eigenvector coalescence. In stark contrast, the solid gray curve shows the same physical dynamics but evaluated at the standard Markovian red-sideband calibration $\Delta^{(0)}_{\rm EP} = -\omega_m$. Tuning to this bare resonance misses the singularity, suppressing the Petermann factor by several orders of magnitude and failing to achieve true mode coalescence. The vertical gray dashed line marks the location of the Markovian exceptional point $G^{(0)}_{\rm EP}/2\pi = 48.75~{\rm kHz}$.}}
\label{fig4}
\end{figure}
As depicted in Fig. (\ref{fig4}), the Petermann factor exhibits a dramatic divergence only when the system is tuned to the memory-renormalized non-Markovian exceptional point, confirming true eigenvector coalescence. Conversely, relying on the standard Markovian calibration bypasses the mathematical singularity, severely suppressing the mode nonorthogonality by several orders of magnitude. This striking contrast visually demonstrates that a precise accounting of environmental memory is necessary to properly locate and exploit the exceptional point.

\section{Spectroscopic signature of non-Markovianity}\label{spec_sec}
\begin{figure}
\centering
\includegraphics[width=\linewidth]{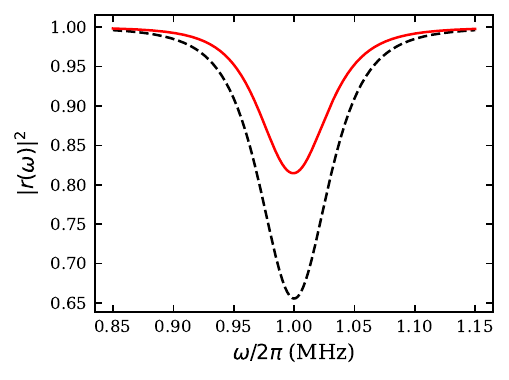}
\caption{\justifying{
Cavity reflection spectrum $|r(\omega)|^2$ as a function of $\omega/2\pi$. The black dashed curve shows the prediction of the Markovian theory evaluated at the Markovian exceptional point $(\Delta^{(0)}_{\rm EP},G^{(0)}_{\rm EP})$, while the red solid curve shows the full non-Markovian response evaluated at the memory-renormalized exceptional point $(\Delta_{\rm EP},G_{\rm EP})$. The model parameters are $\omega_m/2\pi = 1~{\rm MHz}$, $\kappa/2\pi = 0.2~{\rm MHz}$, $\gamma/2\pi = 5~{\rm kHz}$, and $\Omega_c/2\pi = 1~{\rm MHz}$.}}
\label{fig5}
\end{figure}

In order to connect the pseudomode-embedded non-Markovian dynamics with the observable spectrum, it is useful to work in the frequency domain and calculate the cavity reflection function. Starting with the Markovian set of quantum Langevin equations [Eqs. (\ref{a_extended})-(\ref{c_extended})], assuming negligible intrinsic losses, the optical output field satisfies the standard one-sided boundary relation \cite{Gardiner_1985}
\begin{equation}
a_{\rm out}(t)=a_{\rm in}(t)-\sqrt{\kappa} a(t).
\label{io_boundary}
\end{equation}
The quantity $\langle a_{\rm out}(\omega)\rangle/\langle a_{\rm in}(\omega)\rangle $ therefore  describes the optical reflection amplitude of the cavity when a probe field is injected. Introducing Fourier transforms according to $o(\omega)=\int_{-\infty}^{\infty}dt e^{i\omega t}o(t) $, and eliminating the pseudomode, one obtains the frequency-dependent solutions (see Appendix (\ref{appE}))
\begin{eqnarray}
a(\omega)
&=&
\frac{
\sqrt{\kappa}\chi_{b,{\rm eff}}^{-1}(\omega)a_{\rm in}(\omega)
-iG\eta(\omega)\xi(\omega)}{D(\omega)},
\label{aomega_main}\\
b(\omega)
&=&
\frac{\chi_a^{-1}(\omega)\eta(\omega)\xi(\omega)
-iG\sqrt{\kappa}a_{\rm in}(\omega)}{D(\omega)},
\label{bomega_main}
\end{eqnarray}
where the susceptibilities are
\begin{eqnarray}
\chi_a^{-1}(\omega) &=& \frac{\kappa}{2}-i(\omega+\Delta), \\
\chi_{b,{\rm eff}}^{-1}(\omega) &=& \frac{\gamma}{2}-i(\omega-\omega_m) -\frac{g_c^2}{\Omega_c-i\omega},
\end{eqnarray}
while $D(\omega)=\chi_a^{-1}(\omega)\chi_{b,{\rm eff}}^{-1}(\omega)+G^2$, and $\eta(\omega) $ satisfies 
\begin{equation}
|\eta(\omega)|^2
=
\gamma \frac{\omega^2}{\omega^2+\Omega_c^2}.
\label{eta_spectrum_main}
\end{equation}
Substituting Eq. (\ref{aomega_main}) into the boundary relation [Eq. (\ref{io_boundary})] gives
\begin{equation}
a_{\rm out}(\omega) = S_{aa}(\omega) a_{\rm in}(\omega)
+ S_{a\xi}(\omega) \xi(\omega),
\label{aout_decomp_main}
\end{equation}
with the coefficient functions
\begin{eqnarray}
S_{aa}(\omega) &=& 1-\kappa\frac{\chi_{b,{\rm eff}}^{-1}(\omega)}{D(\omega)}, \label{Saa_main} \\
S_{a\xi}(\omega) &=& i\sqrt{\kappa} G \frac{\eta(\omega)}{D(\omega)}. \label{Saxi_main}
\end{eqnarray}
The output-response amplitude therefore reads
\begin{equation}
r(\omega) \equiv \frac{\langle a_{\rm out}(\omega)\rangle}{\langle a_{\rm in}(\omega)\rangle}
= 1-\frac{\kappa}{\chi_a^{-1}(\omega)+G^2\chi_{b,{\rm eff}}(\omega)}.
\label{reflection_main}
\end{equation}
For a one-sided cavity, the quantity $|r(\omega)|^2$ is naturally interpreted as a reflection coefficient. Fig. (\ref{fig5}) shows $|r(\omega)|^2$ as a function of the probe frequency for both the Markovian and non-Markovian descriptions evaluated at their respective exceptional-point parameters. Let us remind the reader that the probe frequency is to be understood as the frequency offset from the control laser. The reflection spectrum furnishes a direct spectroscopic probe of the memory-renormalized dynamics. In both the Markovian and non-Markovian descriptions, the spectrum exhibits a dip near the mechanical resonance $\omega\simeq\omega_m $, corresponding to the familiar optomechanically-induced transparency arising from destructive interference between the intracavity probe field and the field scattered by the mechanics \cite{Aspelmeyer_2014}, although the parameter locking at the exceptional point prevents perfect transparency. The structure of this interference is governed by the mechanical susceptibility entering Eq. (\ref{reflection_main}) so that the non-Markovian modification of the former appears in the reflection spectrum. Because this correction is frequency dependent, it cannot in general be absorbed into a single cooperativity parameter over the full spectrum. Nevertheless, in a narrow frequency window near $\omega\simeq\omega_m$, one may obtain a local estimate by evaluating the dissipative part of the self-energy near the mechanical resonance. This gives the approximate damping renormalization $\gamma_{{\rm eff}} \approx \gamma\left[1-\frac{\Omega_c^2}{\Omega_c^2+\omega_m^2}\right]$, suggesting only locally the replacement
\begin{equation}
C = \frac{4G^2}{\kappa\gamma} \quad \mapsto \quad C_{\rm eff}\approx \frac{4G^2}{\kappa\gamma_{\rm eff}}.
\end{equation}
This estimate is used merely to provide intuition for the change in the transparency dip, and for $\Omega_c \approx \omega_m$, $C_{\rm eff} \approx 2 C$. The spectrum shown in Fig. (\ref{fig5}) (red curve) is computed from the full frequency-dependent response in Eq. (\ref{reflection_main}), and not from the approximate replacement $\gamma\mapsto\gamma_{\rm eff}$. The principal observable consequence of non-Markovianity thus appears as a quantitative modification of the interference condition. Because the mechanical susceptibility is modified by the memory kernel, both the optimal detuning and coupling coordinates and the local interference condition are shifted relative to the Markovian prediction. As a result, the reflection dip predicted by the full non-Markovian theory at its own exceptional-point calibration differs quantitatively from the dip predicted by the Markovian theory at its exceptional point. For our representative parameters $\omega_m/2\pi = 1~{\rm MHz} $, $\kappa/2\pi = 0.2~{\rm MHz} $, $\gamma/2\pi = 5~{\rm kHz} $, and $\Omega_c/2\pi = 1~{\rm MHz} $, the pure Markovian theory and the full non-Markovian theory predict significantly different reflection depths when each is evaluated at its own exceptional-point calibration:
$|r(\omega \simeq \omega_m)|^2_{\rm Markovian} \approx 0.65 $, as opposed to the value $|r(\omega \simeq \omega_m)|^2_{\rm non-Markovian} \approx 0.81 $. As a result, the transparency dip becomes shallower, providing a sensitive spectroscopic signature of the underlying structured mechanical environment.

\section{General structured baths}\label{gen_sec}

Let us briefly discuss to what extent the preceding conclusions on the memory-shifted optomechanical exceptional point depend on the specific spectral density chosen in Eq. (\ref{structured_spectral_density}). The particular form taken in Eq. (\ref{structured_spectral_density}) is analytically convenient because it produces a single-exponential memory kernel [Eq. (\ref{Kt_explicit})] and hence admits an exact Markovian embedding of the drift dynamics using one auxiliary pseudomode. However, the mechanism responsible for the displacement of the exceptional point is more general and can be formulated directly in terms of the bath-induced mechanical self-energy. In this sense, the choice of the spectral density in Eq. (\ref{structured_spectral_density}) merely provides a specific tractable illustration of a universal mechanism.

\vspace{2mm}

For a general structured mechanical environment, the reduced optomechanical characteristic equation is 
\begin{equation}
p(\lambda)=
\left(\lambda-i\Delta+\frac{\kappa}{2}\right)
\left[\lambda+i\omega_m+\frac{\gamma}{2}-\Sigma(\lambda)\right]+G^2=0,
\label{general_self_energy_char}
\end{equation}
where the damping term proportional to $\gamma/2$ denotes the local Markovian contribution that has already been separated from the bath memory kernel, while $\Sigma(\lambda)$ denotes the residual frequency-dependent nonlocal contribution. For the single-pseudomode model considered in Eq. (\ref{structured_spectral_density}), one has $\Sigma(\lambda)=g_c^2/(\Omega_c+\lambda)$. But for a generic reservoir, $\Sigma(\lambda)$ need not be a rational function. Consequently, Eq. (\ref{general_self_energy_char}) is in general a nonlinear spectral equation rather than a polynomial of finite degree. Nevertheless, the condition for a second-order eigenvalue coalescence is still given by Eq. (\ref{double_root_condition_generic}), and thus the existence and location of the optomechanical exceptional point are controlled by the value and local frequency dependence of $\Sigma(\lambda)$ near the relevant window. This observation gives a simple perturbative result for the memory-induced displacement of the exceptional point. If $\Delta_{\rm EP}^{(0)}=-\omega_m$ and $G_{\rm EP}^{(0)}=\frac{\kappa-\gamma}{4}$ are the coordinates of the Markovian exceptional point with coalesced eigenvalue $\lambda_{\rm EP}^{(0)} = -\frac{\kappa+\gamma}{4}-i\omega_m$, then, provided that the bath self-energy is weak and slowly varying near this point, we may evaluate it at the unperturbed exceptional point to obtain
\begin{equation}
\delta \Delta_{\rm EP} = {\rm Im}\left[\Sigma\left(\lambda_{\rm EP}^{(0)}\right)\right],
\quad \quad
\delta G_{\rm EP} = \frac{1}{2}{\rm Re}\left[\Sigma\left(\lambda_{\rm EP}^{(0)}\right)\right],
\label{generic_EP_shift_formula}
\end{equation} up to the leading order. This perturbative result is valid provided
\begin{equation}
\left|\Sigma\left(\lambda_{\rm EP}^{(0)}\right)\right| \ll \omega_m,
\quad \quad 
\left|\Sigma'\left(\lambda_{\rm EP}^{(0)}\right)\right|
\ll 1,
\label{generic_validity_condition}
\end{equation}
so that higher-order eigenvalue dependence of the self-energy can be neglected across the relevant window. It can be verified that substituting the single-pseudomode self-energy of Eq. (\ref{self_energy_model}) reproduces the memory-induced shifts given in Eq. (\ref{shifts_explicit}) in the resolved-sideband and weak-damping regime. 

\vspace{2mm}

It is instructive to compare this with conventional Ohmic, super-Ohmic, and sub-Ohmic mechanical reservoirs. Recalling the relationship between the friction kernel and the bath spectrum in our convention as given in Eq. (\ref{K_int}), we can see at once that $\mathcal{I}(\omega)=\gamma$ produces the local Markovian kernel $K(t)=\gamma\delta(t)$ and therefore represents the Ohmic/Markovian case in the present convention. More generally, one may consider the family of spectral densities
\begin{equation}\label{power_law_spectral_density}
\mathcal{I}_s(\omega) = \gamma \left(\frac{|\omega|}{\omega_m}\right)^s F_c\left(\frac{|\omega|}{\omega_c}\right), \quad \quad s>-1,
\end{equation}
where $F_c$ is a cutoff function that depends on a cutoff-frequency scale $\omega_c$. For this family, the relevant distinction is not the exponent $s$ by itself, but the bath-induced self-energy sampled near the unperturbed optomechanical exceptional point. In the present convention, the Ohmic/Markovian limit corresponds to a frequency-independent effective spectrum $\mathcal{I}(\omega)=\gamma$, or equivalently $s=0$ with a broad cutoff function that is essentially constant near $\omega\simeq\omega_m$. This gives a local memory kernel $K(t)=\gamma\delta(t)$, and therefore reproduces the standard Markovian optomechanical exceptional point. More generally, even for $s=0$, a cutoff close to the mechanical resonance can make the associated self-energy frequency dependent, in which case the dynamics is no longer strictly Markovian. 

\vspace{2mm}

For $s>0$, the reservoir is super-Ohmic in the present convention, and its effect depends on whether the mechanical resonance samples a slowly-varying part of the spectrum or a structured region near a cutoff or crossover scale. Similarly, for $-1<s<0$, corresponding to sub-Ohmic behavior in our convention, the enhanced low-frequency spectral weight can produce longer-lived algebraic memory tails, provided the low-frequency power law persists over the relevant range. In all the cases, the displacement of the exceptional point is controlled not by the exponent alone, but by the value and local frequency dependence of the mechanical self-energy $\Sigma(\lambda)$ near $\lambda_{\rm EP}^{(0)}$. When $\Sigma(\lambda)$ is weak and slowly varying, the leading displacement is given by Eq. (\ref{generic_EP_shift_formula}) subject to the conditions in Eq. (\ref{generic_validity_condition}). When $\Sigma(\lambda)$ varies appreciably over the narrow interval sampled by the coalescing optomechanical modes, stronger memory-induced deviations from the Markovian exceptional point may occur.

\section{Conclusions}\label{conc_sec}
In this work, we showed that non-Markovian mechanical dissipation shifts the exceptional-point location in optomechanical systems. Using a pseudomode-based Markovian embedding of the effective drift dynamics, we derived analytical corrections and parametric conditions for the memory-renormalized exceptional point and quantified its deviation from the standard Markovian prediction. While the induced shifts in detuning and coupling are modest for realistic parameters, their impact on the eigenvector structure is substantial. In particular, the Petermann factor shows that even a small memory-induced displacement strongly suppresses the divergent nonorthogonality when the system is tuned to the Markovian calibration rather than to the true non-Markovian singularity. This highlights that observables sensitive to eigenvector nonorthogonality can provide an effective probe of non-Markovian effects on exceptional points.

\vspace{2mm}

We further showed that these memory-induced modifications appear directly in the optical reflection spectrum through the dressed mechanical susceptibility, producing a measurable change in the optomechanically-induced-transparency dip. Structured mechanical dissipation thus affects both the location of the exceptional point and the associated interference response, the latter manifesting in a form directly measurable in experiments. Our results therefore highlight that environmental memory should be included in the calibration and interpretation of exceptional-point phenomena in optomechanical systems whenever non-Markovian effects are non-negligible. Future work may explore the extension of these ideas to structured phonon reservoirs with frequency-comb-like structure \cite{Qiu_2024_a}, as well as to the non-Markovian collective emission of giant emitters \cite{Qiu_2024_b}.\\

\textbf{Acknowledgements:} We are grateful to the anonymous referees for their constructive comments which have led to a substantial improvement of the paper. A.G. thanks Malay Bandyopadhyay for inspiring discussions. M.B. thanks the Air Force Office of Scientific Research (AFOSR) (FA9550-23-1-0259) for support.

\appendix

\begin{widetext}

\section{Non-Markovian mechanical damping from system-plus-bath approach}\label{appA} 
Considering only the mechanical oscillator $(b,b^\dagger)$, the system-plus-bath Hamiltonian can be taken to be of the form 
\begin{equation} 
H = \Omega_m b^\dagger b + \sum_k \omega_k d_k^\dagger d_k + \sum_k \left(g_k b^\dagger d_k + g_k^* b d_k^\dagger\right), 
\end{equation} 
which appears from the fully-coupled system-plus-bath Hamiltonian upon invoking the rotating-wave approximation \cite{Breuer_Petruccione_2002}. Here $(d_k,d_k^\dagger)$ are the heat bath's degrees of freedom and $\{g_k\}$ are complex-valued system-bath coupling constants. The Heisenberg equations are 
\begin{equation} 
\dot{b} = -i\Omega_m b - i\sum_k g_k d_k, \quad \quad \dot{d}_k = -i\omega_k d_k - i g_k^* b, 
\end{equation} 
which, after solving the bath equation, gives 
\begin{equation} 
d_k(t) = d_k(0)e^{-i\omega_k t} - i g_k^* \int_0^t e^{-i\omega_k(t-\tau)}b(\tau)d\tau. 
\end{equation} 
Substituting this into $\dot{b}$ yields the non-Markovian quantum Langevin equation 
\begin{equation} 
\dot{b} = -i\Omega_m b - \int_0^t K_{\rm micro}(t-\tau)b(\tau)d\tau + F(t), 
\end{equation} where 
\begin{equation} K_{\rm micro}(t) = \sum_k |g_k|^2 e^{-i\omega_k t}, \quad \quad F(t) = -i\sum_k g_k d_k(0)e^{-i\omega_k t}. 
\end{equation} 
In the continuum limit of the bath, we can define the bath spectral function as 
\begin{equation} 
I(\omega) = 2\pi\sum_k |g_k|^2\delta(\omega-\omega_k) \quad \implies \quad K_{\rm micro}(t) = \int_{0}^{\infty} \frac{d\omega}{2\pi} I(\omega)e^{-i\omega t}. 
\end{equation} 
Now in order to derive the noise correlations, let us assume that the bath is initially in a thermal state with the density matrix 
\begin{equation} 
\rho_{\rm B} = \frac{e^{-H_{\rm B}/k_BT}}{Z}. 
\end{equation} 
The bath correlations are thus given by \begin{equation} 
\langle d_k(t)d_l^\dagger(t')\rangle_{\rho_{\rm B}} = \delta_{kl} \left(n(\omega_k)+1\right) e^{-i\omega_k(t-t')}, \quad \quad \langle d_k^\dagger(t)d_l(t')\rangle_{\rho_{\rm B}} = \delta_{kl} n(\omega_k) e^{i\omega_k(t-t')}, \end{equation} 
with thermal Bose factor 
\begin{equation} 
n(\omega) = \frac{1}{e^{\omega/k_B T}-1}. 
\end{equation} 
Putting it all together, some algebra reveals that the noise correlations become 
\begin{equation} \langle F(t)F^\dagger(t')\rangle_{\rho_{\rm B}} = \int_0^\infty \frac{d\omega}{2\pi} I(\omega) (n(\omega)+1)e^{-i\omega(t-t')}, \quad \quad \langle F^\dagger(t)F(t')\rangle_{\rho_{\rm B}} = \int_0^\infty \frac{d\omega}{2\pi} I(\omega) n(\omega) e^{i\omega(t-t')}. 
\end{equation} 
These relations represent the fluctuation-dissipation theorem at finite temperature and match with Eq. (\ref{FDT_mech}) of the main text. 

\vspace{2mm}

\noindent
\textbf{Note:} It is noteworthy that $\Omega_m$ introduced in this appendix differs from $\omega_m$ as $\Omega_m = \omega_m + \Lambda$, where $\Lambda$ has been added here to cancel the frequency renormalization arising from the odd-in-frequency part of the microscopic memory kernel $K_{\rm micro}(t)$, which then allows us to work with the even part of the kernel given in Eq. (\ref{K_int}) of the main text, defined by extending the integration domain to $(-\infty, \infty)$ by introducing the effective two-sided spectral density $\mathcal{I}(\omega)$ used in the main text.

\section{Details of pseudomode mapping}\label{appB}
Starting with the quantum Langevin equation [Eq. (\ref{Lang_b})] of the mechanical mode
\begin{equation}
\dot{b}= -i\omega_m b-iG a -\int_0^t K(t-\tau)b(\tau)d\tau+F(t),
\label{pm_nonmarkov}
\end{equation}
and using Eq. (\ref{Kt_explicit}), one has
\begin{equation}
\dot{b}= -i\omega_m b-iGa -\frac{\gamma}{2}b
+\frac{\gamma\Omega_c}{2}\int_0^t
e^{-\Omega_c (t-\tau)}b(\tau)d\tau+F(t),
\label{pm_split}
\end{equation}
where we have used the symmetric boundary convention
\begin{equation}
\int_0^t \delta(t-\tau)b(\tau)d\tau=\frac{1}{2}b(t), 
\end{equation}
arising when the memory kernel is interpreted as the Fourier transform of an even spectral function defined over the entire real frequency axis. The exponential memory can be generated by introducing and subsequently eliminating an auxiliary pseudomode  $c$. To this end, let us consider the following Markovian quantum Langevin equations:
\begin{eqnarray}
\dot{a} &=& i\Delta a-\frac{\kappa}{2}a-iGb+\sqrt{\kappa} a_{\rm in}(t),\\
\dot{b} &=& -i\omega_m b-\frac{\gamma}{2}b-iGa-g_c c+\sqrt{\gamma} \xi(t),\\
\dot{c} &=& -\Omega_c c- g_c b+\sqrt{2\Omega_c} \xi(t),
\end{eqnarray}
where $c$ is the pseudomode and $\xi(t)$ is a white noise. Solving the equation for $c$ yields
\begin{equation}
c(t)=e^{-\Omega_c t}c(0) -g_c\int_0^t e^{-\Omega_c(t-\tau)}b(\tau)d\tau +\sqrt{2\Omega_c}\int_0^t e^{-\Omega_c(t-\tau)}\xi(\tau)d\tau,
\end{equation}
and substituting this expression into the equation for $b$ leads to
\begin{equation}
\dot{b}=-i\omega_m b-iG a-\frac{\gamma}{2}b+g_c^2\int_0^t e^{-\Omega_c(t-\tau)}b(\tau)d\tau
+F_{\rm eff}(t).
\end{equation}
Comparing with Eq. (\ref{pm_split}) forces us to fix
\begin{equation}
g_c^2 = \frac{\gamma\Omega_c}{2}.
\end{equation}
Now the effective noise appearing in the mechanical equation is
\begin{equation}
F_{\rm eff}(t) = \sqrt{\gamma} \xi(t) -g_c\sqrt{2\Omega_c}
\int_0^t e^{-\Omega_c(t-\tau)}\xi(\tau)d\tau .
\end{equation}
The term proportional to $c(0)$ represents an initial transient that decays as $e^{-\Omega_c t}$ and can therefore be neglected when focusing on the long-time stationary regime. From the white-noise correlations
\begin{equation}
\langle \xi(t)\xi^\dagger(t') \rangle = (n_{\rm th}+1)\delta(t-t'), 
\quad \quad
\langle \xi^\dagger(t)\xi(t') \rangle = n_{\rm th} \delta(t-t'),
\end{equation}
one obtains
\begin{eqnarray}
\langle F_{\rm eff}(t)F_{\rm eff}^\dagger(t')\rangle
&=& (n_{\rm th}+1)
\int_{-\infty}^{\infty}\frac{d\omega}{2\pi} 
\gamma\frac{\omega^2}{\omega^2+\Omega_c^2}
e^{-i\omega(t-t')}, \\
\langle F_{\rm eff}^\dagger(t)F_{\rm eff}(t')\rangle
&=& n_{\rm th}
\int_{-\infty}^{\infty}\frac{d\omega}{2\pi} 
\gamma\frac{\omega^2}{\omega^2+\Omega_c^2}
e^{i\omega(t-t')}.
\end{eqnarray}
In the long-time stationary regime, the effective noise spectrum is therefore governed by
\begin{equation}
\mathcal{I}(\omega)=\gamma\frac{\omega^2}{\omega^2+\Omega_c^2}.
\end{equation}
The pseudomode construction thus reproduces the target exponential memory kernel up to the decaying initial transient discussed above, and generates effective noise correlations associated with the even spectral function $\mathcal{I}(\omega)$, provided that the thermal occupation is approximately frequency independent over the relevant bandwidth, i.e., $n(\omega)\simeq n_{\rm th}$.

\section{Schur-complement reduction and effective mechanical self-energy}\label{appC}
Starting from the $3 \times 3$ drift matrix $M$ as given in Eq. (\ref{general_drift_matrix}), the characteristic equation is $\det(\lambda \mathbb{I} - M) = 0$. In order to obtain an effective $2 \times 2$ matrix for the optomechanical block, it is convenient to write
\begin{equation}
\lambda \mathbb{I} - M =
\begin{pmatrix}
\mathcal{A}(\lambda) & \mathcal{B} \\
\mathcal{C} & \mathcal{D}(\lambda)
\end{pmatrix},
\end{equation}
with
\begin{equation}
\mathcal{A}(\lambda)=
\begin{pmatrix}
\lambda - i\Delta + \frac{\kappa}{2} & iG \\
iG & \lambda + i\omega_m + \frac{\gamma}{2}
\end{pmatrix},
\quad
\mathcal{B}=
\begin{pmatrix}
0\\
g_c
\end{pmatrix},
\quad
\mathcal{C}=
\begin{pmatrix}
0 & g_c
\end{pmatrix},
\quad
\mathcal{D}(\lambda)=\lambda + \Omega_c.
\end{equation}
Now using the Schur-complement identity that goes as
\begin{equation}
\det(\lambda \mathbb{I} - M) = \det \mathcal{D}(\lambda) \det \left[\mathcal{A}(\lambda)-\mathcal{B} \mathcal{D}(\lambda)^{-1} \mathcal{C}\right],
\end{equation}
valid for $\mathcal{D}(\lambda)\neq 0$, one obtains
\begin{equation}
\det(\lambda \mathbb{I} - M) = (\lambda+\Omega_c)
\det \left[\mathcal{A}(\lambda)-\mathcal{B} \mathcal{D}(\lambda)^{-1} \mathcal{C}\right].
\end{equation}
Since
\begin{equation}
\mathcal{B} \mathcal{D}(\lambda)^{-1} \mathcal{C}
=
\begin{pmatrix}
0 & 0\\
0 & \dfrac{g_c^2}{\Omega_c+\lambda}
\end{pmatrix},
\end{equation}
the Schur complement becomes
\begin{equation}
\mathcal{A}(\lambda)-\mathcal{B} \mathcal{D}(\lambda)^{-1} \mathcal{C} =
\begin{pmatrix}
\lambda - i\Delta + \frac{\kappa}{2} & iG \\
iG &
\lambda + i\omega_m + \frac{\gamma}{2}
- \dfrac{g_c^2}{\Omega_c+\lambda}
\end{pmatrix}.
\end{equation}
Up to the prefactor $(\lambda+\Omega_c)$ associated with the eliminated pseudomode, therefore, the optomechanical eigenvalue problem is governed by an effective $2\times 2$ matrix
\begin{equation}
M_{{\rm eff}}(\lambda)=
\begin{pmatrix}
i\Delta-\frac{\kappa}{2} & -iG \\
-iG &
-\left(i\omega_m+\frac{\gamma}{2}\right)+\Sigma(\lambda)
\end{pmatrix},
\quad \quad 
\Sigma(\lambda)=\frac{g_c^2}{\Omega_c+\lambda},
\end{equation}
such that the reduced characteristic equation is
\begin{equation}
\det \bigl(\lambda \mathbb{I}-M_{\rm eff}(\lambda)\bigr)=0.
\end{equation}
This reduction is exact (assuming $\lambda \neq -\Omega_c$) and shows that the effect of the eliminated pseudomode is to generate a frequency-dependent mechanical self-energy $\Sigma(\lambda)$ that encodes the non-Markovian memory of the structured reservoir.

\section{Calculation of non-Markovian exceptional points}\label{appD}
The characteristic polynomial $p(\lambda)$ of the drift matrix $M$ [Eq. (\ref{general_drift_matrix})] can be compactly expressed as 
\begin{equation}\label{p_lambda_compact}
p(\lambda)= (\lambda-i\Delta+\tfrac{\kappa}{2})h(\lambda)
+g(\lambda)G^2,
\end{equation} where $g(\lambda)$ and $h(\lambda)$ are defined in Eqs. (\ref{g_def}) and (\ref{h_def}). A second-order exceptional point implies the presence of a double root of the characteristic polynomial, requiring
\begin{equation}\label{EP_condition}
p(\lambda_{\rm EP})=0,
\quad \quad
p'(\lambda_{\rm EP})=0,
\quad \quad
p''(\lambda_{\rm EP}) \neq 0,
\end{equation}
and if the matrix becomes defective, this second-order degeneracy is an exceptional point. Differentiating Eq. (\ref{p_lambda_compact}) gives
\begin{equation}
p'(\lambda) = h(\lambda) + (\lambda-i\Delta+\tfrac{\kappa}{2})h'(\lambda) + G^2,
\end{equation}
since $g'(\lambda)=1$. Using $h(\lambda)=g(\lambda)f(\lambda)-g_c^2$, one finds
\begin{equation}
h'(\lambda)=g'(\lambda)f(\lambda)+g(\lambda)f'(\lambda)
=f(\lambda)+g(\lambda),
\end{equation}
because $f'(\lambda)=1$. At the exceptional point, the conditions $p(\lambda_{\rm EP})=0$ and $p'(\lambda_{\rm EP})=0$ therefore give the pair of equations
\begin{eqnarray}
(\lambda_{\rm EP}-i\Delta+\tfrac{\kappa}{2})h(\lambda_{\rm EP})
+g(\lambda_{\rm EP})G_{\rm EP}^2 &=&0, \label{EP_eq1} \\
h(\lambda_{\rm EP})
+(\lambda_{\rm EP}-i\Delta+\tfrac{\kappa}{2})
h'(\lambda_{\rm EP})
+G_{\rm EP}^2 &=&0 . \label{EP_eq2}
\end{eqnarray}
These two relations are linear in the two unknowns
$(\lambda_{\rm EP}-i\Delta+\kappa/2)$ and $G_{\rm EP}^2$. Solving Eq. (\ref{EP_eq1}) for $G_{\rm EP}^2$ gives
\begin{equation}
G_{\rm EP}^2 = -\frac{(\lambda_{\rm EP}-i\Delta+\tfrac{\kappa}{2}) h(\lambda_{\rm EP})}{g(\lambda_{\rm EP})}.
\end{equation}
Substituting this into Eq. (\ref{EP_eq2}) yields
\begin{equation}
h(\lambda_{\rm EP}) + (\lambda_{\rm EP}-i\Delta+\tfrac{\kappa}{2}) \left[h'(\lambda_{\rm EP})
-\frac{h(\lambda_{\rm EP})}{g(\lambda_{\rm EP})}
\right]=0.
\end{equation}
Now solving for $(\lambda_{\rm EP}-i\Delta+\kappa/2)$ gives
\begin{equation}\label{B_solution}
\lambda_{\rm EP}-i\Delta_{\rm EP}+\frac{\kappa}{2} =
-\frac{g(\lambda_{\rm EP})h(\lambda_{\rm EP})}{g(\lambda_{\rm EP})^2+g_c^2}.
\end{equation}
Using this result in Eq. (\ref{EP_eq1}) yields the coupling strength at the exceptional point to be
\begin{equation}\label{G_EP_formula}
G_{\rm EP}^2 = \frac{h(\lambda_{\rm EP})^2}{g(\lambda_{\rm EP})^2+g_c^2},
\end{equation} matching with Eq. (\ref{G_EP_general}) of the main text. Eq. (\ref{B_solution}) now gives the expression for the exceptional-point detuning as
\begin{equation}\label{Delta_EP_formula}
\Delta_{\rm EP}(\lambda) = -i\left[
\lambda+\frac{\kappa}{2} + \frac{g(\lambda)h(\lambda)}
{g(\lambda)^2+g_c^2}\right],
\end{equation}
matching with Eq. (\ref{Delta_EP_general}). One can now check that $p''(\lambda_{\rm EP})\neq 0$, implying that the algebraic multiplicity is two, not three. Finally, let us note that if $\lambda_{\rm EP}$ denotes the double root of the cubic, the characteristic polynomial may be written as
\begin{equation}
p(\lambda)=(\lambda-\lambda_{\rm EP})^2(\lambda-\lambda_3).
\end{equation}
Using the well-known Vi\`ete relation for the sum of the roots yields
\begin{equation}
2\lambda_{\rm EP}+\lambda_3 = -\left[
\Omega_c+\frac{\gamma}{2}+\frac{\kappa}{2} +i(\omega_m-\Delta)\right],
\end{equation}
and therefore
\begin{equation}
\lambda_3 = -\left[\Omega_c+\frac{\gamma}{2}+\frac{\kappa}{2}+i(\omega_m-\Delta)\right] -2\lambda_{\rm EP}.
\end{equation}

\section{Calculation details of the input-output formalism of the pseudomode-embedded system}\label{appE}
Starting with Fourier transforms in the convention
\begin{equation}
o(\omega)=\int_{-\infty}^{\infty}dt e^{i\omega t}o(t),
\quad \quad
o(t)=\int_{-\infty}^{\infty}\frac{d\omega}{2\pi} e^{-i\omega t}o(\omega),
\label{app_io_fourier}
\end{equation}
i.e.,
\begin{equation}
\int_{-\infty}^{\infty}dt e^{i\omega t}\dot{o}(t) =
-i\omega o(\omega),
\label{app_io_derivative}
\end{equation}
the quantum Langevin equations [Eqs. (\ref{a_extended})-(\ref{c_extended})] read
\begin{eqnarray}
-i\omega a(\omega) &=& \left(i\Delta-\frac{\kappa}{2}\right)a(\omega)-iG b(\omega)+\sqrt{\kappa} a_{\rm in}(\omega), \\
-i\omega b(\omega) &=& -\left(i\omega_m+\frac{\gamma}{2}\right)b(\omega)-iG a(\omega)-g_c c(\omega)+\sqrt{\gamma} \xi(\omega), \\
-i\omega c(\omega) &=& -\Omega_c c(\omega)-g_c b(\omega)+\sqrt{2\Omega_c} \xi(\omega).
\end{eqnarray}
Rearranging, one obtains
\begin{eqnarray}
\left[\frac{\kappa}{2}-i(\omega+\Delta)\right]a(\omega)+iG b(\omega) &=& \sqrt{\kappa} a_{\rm in}(\omega), \label{app_io_a_freq_raw}\\
iG a(\omega)+\left[\frac{\gamma}{2}-i(\omega-\omega_m)\right]b(\omega)+g_c c(\omega) &=& \sqrt{\gamma} \xi(\omega), \label{app_io_b_freq_raw}\\
g_c b(\omega)+\left(\Omega_c-i\omega\right)c(\omega)
&=& \sqrt{2\Omega_c} \xi(\omega). \label{app_io_c_freq_raw}
\end{eqnarray}
It is therefore natural to define the bare susceptibilities
\begin{eqnarray}
\chi_a^{-1}(\omega) &=& \frac{\kappa}{2}-i(\omega+\Delta), \label{app_io_chia}\\
\chi_b^{-1}(\omega) &=& \frac{\gamma}{2}-i(\omega-\omega_m), \label{app_io_chib}\\
\chi_c^{-1}(\omega) &=& \Omega_c-i\omega, \label{app_io_chic}
\end{eqnarray}
so that the coupled equations take the compact form
\begin{eqnarray}
\chi_a^{-1}(\omega)a(\omega)+iG b(\omega) &=&
\sqrt{\kappa} a_{\rm in}(\omega), \label{app_io_a_freq}\\
iG a(\omega)+\chi_b^{-1}(\omega)b(\omega)+g_c c(\omega) &=& \sqrt{\gamma} \xi(\omega), \label{app_io_b_freq}\\
g_c b(\omega)+\chi_c^{-1}(\omega)c(\omega)
&=& \sqrt{2\Omega_c} \xi(\omega). \label{app_io_c_freq}
\end{eqnarray}
Now from Eq. (\ref{app_io_c_freq}), the pseudomode amplitude is
\begin{equation}
c(\omega) = \chi_c(\omega)\left[\sqrt{2\Omega_c} \xi(\omega)-g_c b(\omega)\right].
\label{app_io_csolution}
\end{equation}
Substituting this into Eq. (\ref{app_io_b_freq}) yields
\begin{equation}
iG a(\omega)+\chi_b^{-1}(\omega)b(\omega) + g_c\chi_c(\omega)\left[\sqrt{2\Omega_c} \xi(\omega)-g_c b(\omega)\right] = \sqrt{\gamma} \xi(\omega),
\end{equation}
i.e.,
\begin{equation}
iG a(\omega) + \left[\chi_b^{-1}(\omega)-g_c^2\chi_c(\omega)\right]b(\omega) = \left[\sqrt{\gamma}-g_c\sqrt{2\Omega_c} \chi_c(\omega)\right]\xi(\omega).
\label{app_io_reduced_pre}
\end{equation}
This identifies the dressed mechanical susceptibility
\begin{eqnarray}
\chi_{b,{\rm eff}}^{-1}(\omega) &=& \chi_b^{-1}(\omega)-g_c^2\chi_c(\omega) \nonumber \\
&=& \chi_b^{-1}(\omega)-\frac{g_c^2}{\Omega_c-i\omega},
\label{app_io_chimeff}
\end{eqnarray}
and the transfer function
\begin{equation}
\eta(\omega) = \sqrt{\gamma}-g_c\sqrt{2\Omega_c} \chi_c(\omega).
\label{app_io_eta}
\end{equation}
The reduced cavity-mechanical equations are therefore
\begin{eqnarray}
\chi_a^{-1}(\omega)a(\omega)+iG b(\omega) &=&
\sqrt{\kappa} a_{\rm in}(\omega), \label{app_io_red1}\\
iG a(\omega)+\chi_{b,{\rm eff}}^{-1}(\omega)b(\omega)
&=& \eta(\omega)\xi(\omega). \label{app_io_red2}
\end{eqnarray}
Using the relation $g_c^2=\frac{\gamma\Omega_c}{2}$, Eq. (\ref{app_io_eta}) simplifies to
\begin{eqnarray}
\eta(\omega) &=& \sqrt{\gamma}-\frac{g_c\sqrt{2\Omega_c}}{\Omega_c-i\omega}
= \sqrt{\gamma}-\frac{\sqrt{\gamma\Omega_c/2} \sqrt{2\Omega_c}}{\Omega_c-i\omega}
\nonumber\\
&=& \sqrt{\gamma}\left(1-\frac{\Omega_c}{\Omega_c-i\omega}\right) = -\sqrt{\gamma} \frac{i\omega}{\Omega_c-i\omega},
\label{app_io_eta_simplified}
\end{eqnarray}
thereby giving
\begin{equation}
|\eta(\omega)|^2 = \gamma \frac{\omega^2}{\omega^2+\Omega_c^2}.
\label{app_io_eta_modsq}
\end{equation}
To solve the reduced system of Eqs. (\ref{app_io_red1}) and (\ref{app_io_red2}), let us write them in the matrix form as
\begin{equation}
\begin{pmatrix}
\chi_a^{-1}(\omega) & iG \\
iG & \chi_{b,{\rm eff}}^{-1}(\omega)
\end{pmatrix}
\begin{pmatrix}
a(\omega)\\
b(\omega)
\end{pmatrix}
=
\begin{pmatrix}
\sqrt{\kappa} a_{\rm in}(\omega)\\
\eta(\omega) \xi(\omega)
\end{pmatrix}.
\label{app_io_matrix}
\end{equation}
The determinant of the coefficient matrix is
\begin{eqnarray}
D(\omega)
&=& \chi_a^{-1}(\omega)\chi_{b,{\rm eff}}^{-1}(\omega)-(iG)^2 \nonumber\\
&=& \chi_a^{-1}(\omega)\chi_{b,{\rm eff}}^{-1}(\omega)+G^2.
\label{app_io_D}
\end{eqnarray}
Inverting the  $2\times2$ matrix gives
\begin{equation}
\begin{pmatrix}
a(\omega)\\
b(\omega)
\end{pmatrix}
=
\frac{1}{D(\omega)}
\begin{pmatrix}
\chi_{b,{\rm eff}}^{-1}(\omega) & -iG \\
-iG & \chi_a^{-1}(\omega)
\end{pmatrix}
\begin{pmatrix}
\sqrt{\kappa} a_{\rm in}(\omega)\\
\eta(\omega) \xi(\omega)
\end{pmatrix}.
\end{equation}
Hence
\begin{eqnarray}
a(\omega) &=& \frac{\sqrt{\kappa} \chi_{b,{\rm eff}}^{-1}(\omega) a_{\rm in}(\omega)-iG \eta(\omega) \xi(\omega)}{D(\omega)},
\label{app_io_a_solution}
\\
b(\omega) &=& \frac{\chi_a^{-1}(\omega)\eta(\omega) \xi(\omega) -iG\sqrt{\kappa} a_{\rm in}(\omega)}{D(\omega)},
\label{app_io_b_solution}
\end{eqnarray}
which match with Eqs. (\ref{aomega_main}) and (\ref{bomega_main}) of the main text. Finally, substituting Eq. (\ref{app_io_a_solution}) into the input-output relation of Eq. (\ref{io_boundary}) gives
\begin{eqnarray}
a_{\rm out}(\omega) &=& a_{\rm in}(\omega)-\sqrt{\kappa} a(\omega) \nonumber\\
&=& a_{\rm in}(\omega)-\sqrt{\kappa} \frac{\sqrt{\kappa} \chi_{b,{\rm eff}}^{-1}(\omega) a_{\rm in}(\omega)-iG \eta(\omega) \xi(\omega)}{D(\omega)}.
\end{eqnarray}
This can be rearranged to the form
\begin{equation}
a_{\rm out}(\omega) = S_{aa}(\omega) a_{\rm in}(\omega) + S_{a\xi}(\omega) \xi(\omega),
\label{app_io_aout}
\end{equation}
with
\begin{eqnarray}
S_{aa}(\omega) &=& 1-\kappa\frac{\chi_{b,{\rm eff}}^{-1}(\omega)}{D(\omega)},
\label{app_io_Saa}
\\
S_{a\xi}(\omega) &=&
i\sqrt{\kappa} G \frac{\eta(\omega)}{D(\omega)}.
\label{app_io_Saxi}
\end{eqnarray}
In the absence of a coherent drive on the mechanical bath, one has $\langle \xi(\omega)\rangle =0$,
so the optical output response is
\begin{equation}
r(\omega) \equiv \frac{\langle a_{\rm out}(\omega)\rangle}{\langle a_{\rm in}(\omega)\rangle}
= S_{aa}(\omega).
\end{equation}
Using Eq. (\ref{app_io_D}), this may also be written as
\begin{eqnarray}
r(\omega) &=& 1-\kappa\frac{\chi_{b,{\rm eff}}^{-1}(\omega)}{\chi_a^{-1}(\omega)\chi_{b,{\rm eff}}^{-1}(\omega)+G^2} \nonumber\\
&=& 1-\frac{\kappa}{\chi_a^{-1}(\omega)+G^2\chi_{b,{\rm eff}}(\omega)}.
\label{app_io_reflection}
\end{eqnarray}
This matches with Eq. (\ref{reflection_main}) of the main text. 

\end{widetext}

\end{document}